\documentclass[times,twocolumn,final]{elsarticle}

\usepackage{medima}
\usepackage{framed,multirow}

\usepackage{amssymb}
\usepackage{latexsym}

\usepackage{subcaption}  % <----

\usepackage{url}
\usepackage{xcolor}

\usepackage{hyperref}

\definecolor{newcolor}{rgb}{.8,.349,.1}

\journal{Medical Image Analysis}

\begin{document}

\verso{Numan Saeed \textit{et~al.}}

\begin{frontmatter}

\title{MGMT promoter methylation status prediction using MRI scans? An extensive experimental evaluation of deep learning models}%
% \tnotetext[tnote1]{This is an example for title footnote coding.}

\author[1]{Numan \snm{Saeed}\corref{cor1}}
\cortext[cor1]{Corresponding author: 
    Numan.Saeed@mbzuai.ac.ae}
\author[1]{Muhammad \snm{Ridzuan}}
\author[1]{Hussain \snm{Alasmawi}}
\author[2]{Ikboljon \snm{Sobirov}}
\author[2]{Mohammad \snm{Yaqub}}

\address[1]{Department of Machine Learning, Mohamed bin Zayed University of Artificial Intelligence, Abu Dhabi, United Arab Emirates}
\address[2]{Department of Computer Vision, Mohamed bin Zayed University of Artificial Intelligence, Abu Dhabi, United Arab Emirates}

% \received{1 May 2013}
% \finalform{10 May 2013}
% \accepted{13 May 2013}
% \availableonline{15 May 2013}
% \communicated{S. Sarkar}

\begin{abstract}
\textit{The number of studies on deep learning for medical diagnosis is expanding, and these systems are often claimed to outperform clinicians. However, only a few systems have shown medical efficacy. From this perspective, we examine a wide range of deep learning algorithms for the assessment of glioblastoma - a common brain tumor in older adults that is lethal. Surgery, chemotherapy, and radiation are the standard treatments for glioblastoma patients. The methylation status of the MGMT promoter, a specific genetic sequence found in the tumor, affects chemotherapy's effectiveness. MGMT promoter methylation improves chemotherapy response and survival in several cancers. MGMT promoter methylation is determined by a tumor tissue biopsy, which is then genetically tested. This lengthy and invasive procedure increases the risk of infection and other complications. Thus, researchers have used deep learning models to examine the tumor from brain MRI scans to determine the MGMT promoter's methylation state. We employ deep learning models and one of the largest public MRI datasets of 585 participants to predict the methylation status of the MGMT promoter in glioblastoma tumors using MRI scans. We test these models using Grad-CAM, occlusion sensitivity, feature visualizations, and training loss landscapes. Our results show no correlation between these two, indicating that external cohort data should be used to verify these models' performance to assure the accuracy and reliability of deep learning systems in cancer diagnosis.}
\end{abstract}


\end{frontmatter}

\section{Introduction}

Glioblastoma multiforme (GBM) is one of the most aggressive brain tumors, with a dismal survival rate and few treatments. Each year, the projected number of GBM diagnoses and fatalities in the United States is over 13,000 and over 10,000, respectively \citep{ostrom2019primary}.
According to the World Health Organization, GBM is considered the most severe brain cancer (grade 5). After the tumor is surgically removed, chemotherapy and radiation are typically the standard courses of treatment.
However, there are significant risks associated with radiotherapy since radiation has the potential to harm both malignant and healthy cells.  Conversely, chemotherapy kills cancer cells by inducing apoptosis and stopping their reproduction by attaching a chemical to the guanine DNA.  O$^6$-methylguanine DNA methyltransferase (MGMT) is a recognized factor in the ineffectiveness of chemotherapy. An inadequate response to temozolomide (TMZ) is caused by the DNA repair enzyme of MGMT, which lessens the impact of alkylating chemotherapy drugs on tumor cells. The methylation status of the MGMT promoter determines its working mechanism \citep{esteller2000inactivation, brandes2009recurrence}. Patients with methylated MGMT promoters have a median survival rate of 21.7 months, as opposed to 15.3 months for those with unmethylated MGMT promoters~\citep{hegi2005mgmt}. Chemotherapy treatment may be successful if the promoter region is methylated since this alters the transcription of the enzyme. The MGMT promoter methylation status has thus evolved into a prognostic indicator and a predictor of chemotherapy response. Typically, invasive methods like biopsy or open surgical resection are used to gather information on gliomas' molecular and genetic changes. However, these methods take time and effort and raise the danger of infection.

Radiogenomics, often known as imaging genomics, is a relatively new approach that establishes links between the genetic makeup of cancer and its imaging properties. The ability of magnetic resonance imaging (MRI) to predict the genetics of brain cancer is one area in particular that has drawn much interest. Previous research relied on radiologists manually extracting imaging information, which is a labor-intensive and subjective approach. Automatic feature extraction \citep{zhou2014learning}, object detection \citep{smirnov2014comparison}, and image classification \citep{krizhevsky2017imagenet} using deep learning techniques outperform manual methods.
 
Numerous studies \citep{chang2018deep, yogananda2020mri} have demonstrated the effectiveness of deep learning techniques in predicting the MGMT promoter methylation status using imaging data. Other studies, however, have found that MRI scans cannot predict MGMT promoter status \citep{egana2020methylation} and are not a reliable predictive indicator for the TMZ response \citep{han2018mri, mikkelsen2020mgmt}.

This study questions whether data from brain MR scans can (or cannot) be reliably used to predict MGMT promoter status. We examine state-of-the-art deep-learning models' capability to classify the MGMT promoter's methylation status from MRI scans. We use one of the largest MRI datasets for this investigation, where several MRI modalities are captured for each patient. 

This work is an extension of our previous work~\citep{saeed2022possible} at the Medical Imaging with Deep Learning (MIDL) 2022 conference, where we studied the relation between MR images and MGMT methylation. This paper extensively extends on \citep{saeed2022possible} with the following main contributions:
\begin{itemize}
    \item All experiments reported in the MIDL paper are complemented with more detailed explanations.
    \item We add Task 1 dataset that is provided for the segmentation task. Preprocessing steps are all provided for this dataset, and there are no inconsistencies or inaccuracies in them. More details are provided in Section~\ref{section:dataset}.
    \item We conduct the additional evaluation by introducing more models, modalities, and cross-validation. We use DenseNets, EfficientNets, and transformer models apart from ResNets in various combinations with T1wCE, T2, and FLAIR modalities using a 5-fold cross-validation.
    \item We add loss landscapes using feature normalization and confirm that the binary cross-entropy (BCE) loss is indeed getting plateaued at around 0.7, indicating randomness in the model prediction. 
    \item We thoroughly perform model interpretability analyses to understand the inner dynamics of the models using Grad-CAM and occlusion sensitivity as described in Section~\ref{section:discussion}.
    \item Furthermore, we extend the feature visualization analyses with PCA and t-SNE for MGMT prediction and compare them against another medical dataset for the binary classification of lung nodules. Our findings show that the ResNet-10 model cannot differentiate between the classes for MGMT prediction but, conversely is able to perform the classification on the lung nodule dataset easily. 
    \item Finally, we add a section for recommendations on the development and evaluation processes of deep learning models in cancer diagnosis.  
\end{itemize}

\subsection{Related work}

Numerous attempts using deep learning techniques have recently claimed promising outcomes in classifying the MGMT promoter methylation status through MRI scans. Using T2, FLAIR, and T1-weighted pre- and post-contrast MRI scans, \cite{chang2018deep} trained a convolutional neural network (CNN) with residual connections to classify the methylation status and obtained a mean accuracy score of 83\% on 5-fold cross-validation. The data was collected from The Cancer Imaging Archives (TCIA) \citep{clark2013cancer} and The Cancer Genome Atlas (TCGA) \citep{tomczak2015cancer}, where both low and high-grade gliomas are considered. In a different study by \cite{yogananda2020mri}, they developed an MGMT-net, T2WI-only network based on 3D-Dense-UNets for MGMT methylation status determination in addition to tumor segmentation. They stated that the mean cross-validation accuracy across three folds was 94.73\%, while sensitivity and specificity scores were 96.31\% and 91.66\%, respectively. \cite{korfiatis2017residual} utilized ResNet-50 and showed an accuracy of 94.90\% on a testing set, similar to \cite{chang2018deep}, but utilizing complete T2 scans done at the Mayo Clinic with documented MGMT methylation information. \cite{mun2022multi} developed a multi-modal late fusion 3D classification network that can be modified to incorporate radiomics characteristics or other external features to classify MGMT promoter methylation status using 3D MRI scans utilizing multiple modalities (T1w, T1wCE, T2w, FLAIR). The AUC score for their top-performing classification model was 0.698. 

\begin{figure*}[!t]
\centering
\begin{tabular}{ccc}
    {\includegraphics[width=0.545\columnwidth]{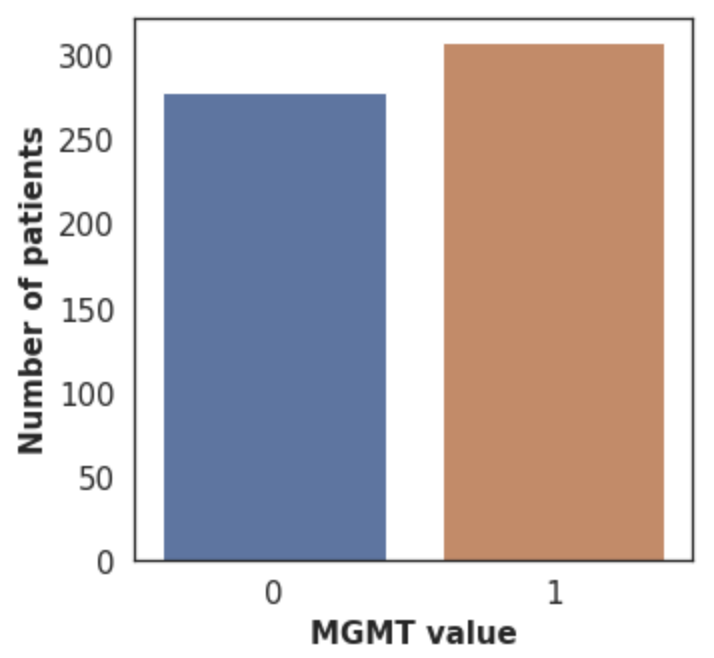}}&

    {\includegraphics[width=0.6\columnwidth]{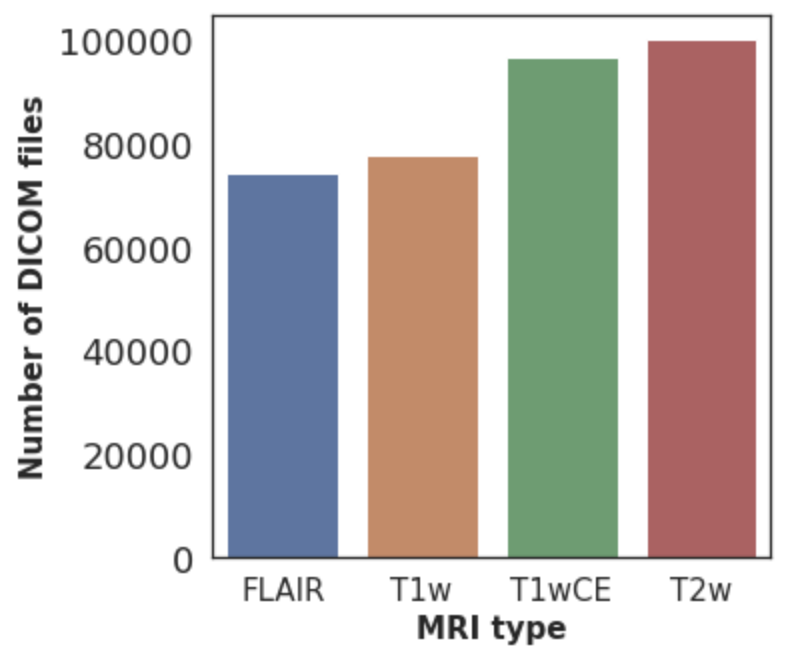}}&

    {\includegraphics[width=0.72\columnwidth]{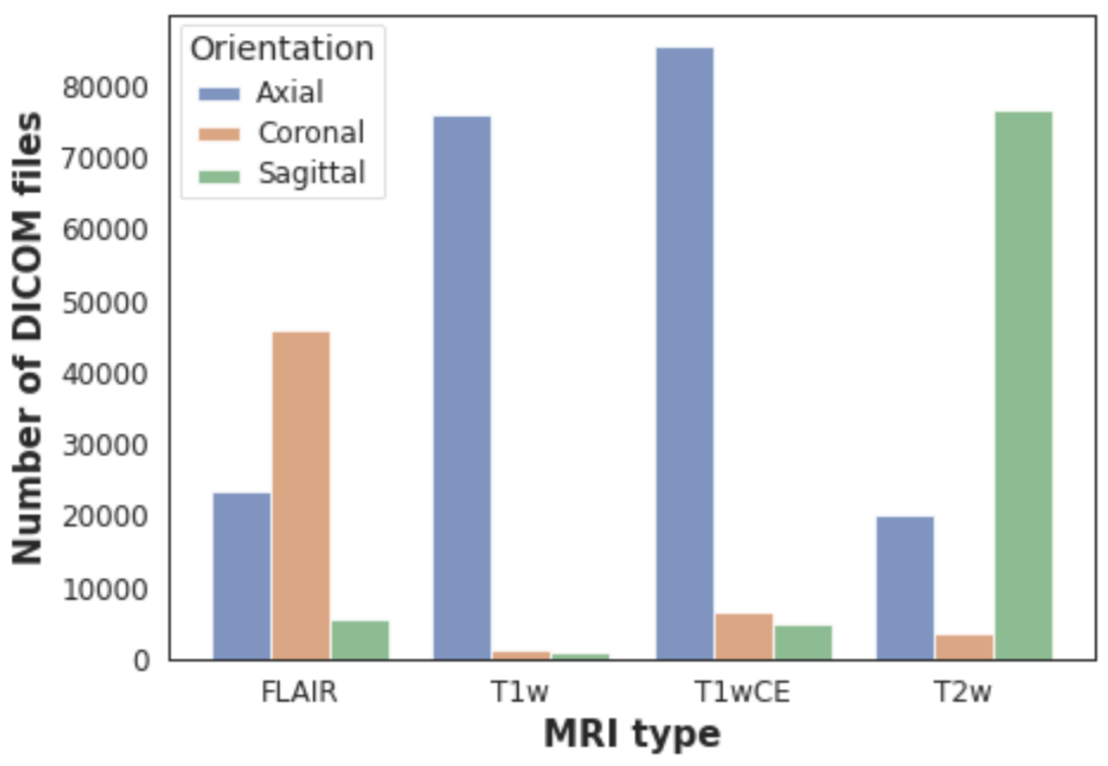}}
    \\
    (a)&(b)&(c)
\end{tabular}
\caption{Dataset distribution based on different features. (a) shows that the dataset contains an equal number of patients for the two classes. (b) shows the difference in the number of slices of different MRI types. (c) shows by further drilling down the orientation distribution for each MRI type.}
\label{fig:MRI-dist}
\end{figure*}

Despite the success of traditional CNNs, recent findings raise some doubts about the predictive power of MGMT methylation from MRI scans and the viability of resolving this issue using deep learning approaches. For example, \cite{han2018mri} employed a recurrent CNN model to predict the MGMT methylation from MRI scans and could not achieve more than 62\% accuracy on the test dataset with precision and recall of 67\% each. This test was conducted using molecular genetic data and MRI scans of T1, T2, and FLAIR modalities from TCIA and TCGA. Another example are the model implementations of the participants in the Brain Tumor Radiogenomic Classification challenge \citep{baid2021rsna} where participants used the most comprehensive dataset for this task; the winning solution could not achieve more than a 0.62 AUC score. In a recent study \citep{emchinov2022deep}, authors used the same comprehensive dataset where they proposed a method based on the similarity of four neural networks; one 3D-ResNet-34 neural network for each type of scan. After testing several loss functions, they found that the binary-cross entropy had the best convergence among other loss functions achieving an average AUC validation standing at 0.5994. Another recent study \citep{qu2022attentive} employed the same dataset and proposed a novel deep neural network design that combines three performance enhancers, including a light attention mechanism, a separable embedding module, and a model-wise shortcut method, and achieved an accuracy of only 63.71\%. These deep-learning-based studies showed that such models could not detect the MGMT methylation from MRI scans, contrary to the previously cited studies. In a recent clinical study, epigenetic silencing of the MGMT promoter could not predict the response to TMZ in a cohort of 334 patients with glioblastoma or high-grade glioma \citep{egana2020methylation}. According to the study, there was no connection between the methylation of MGMT promoters and overall survival rate. In \cite{mikkelsen2020mgmt}, authors looked into the clinical relationships between this prognostic biomarker and several radiological and histological characteristics in patients with dehydrogenase (IDH) wild-type glioblastomas. No connections were found between the examined factors, including the MRI images' characteristics, overall survival, and MGMT status. According to the authors, the methylation status of MGMT cannot be non-invasively predicted from MRI characteristics.

\section{Dataset}
\label{section:dataset}
The Medical Image Computing and Computer Assisted Intervention Society (MICCAI) and the Radiological Society of North America (RSNA) have partnered to enhance patients' glioblastoma diagnosis and treatment plan. One of the top aims of contemporary medicine is to reduce glioblastoma categorization procedures. In light of this, RSNA and MICCAI have created a global competition called Brain Tumor Radiogenomic Classification \citep{baid2021rsna}. The primary objective of the competition is to create intelligent solutions to identify the MGMT promoter status using MRI scans of glioblastoma patients. The dataset utilized in this work comes from the Brain Tumor Radiogenomic Classification challenge. It consists of a collection of de-identified public and private datasets from the TCIA \citep{clark2013cancer}, a public collection of TCGA-GBM, the ACRIN-FMISO-Brain collection (ACRIN 6684) \citep{gerstner2016acrin}, and others. The dataset was used for two tasks in the 2021 competition; 1) Segmentation Task\footnote{http://braintumorsegmentation.org/} and 2) Radiogenomic Classification Task\footnote{https://www.kaggle.com/c/rsna-miccai-brain-tumor-radiogenomic-classification}. The dataset in the classification task is smaller than that of the segmentation task as it consists only of MRIs of patients from the segmentation task for whom the MGMT-promoter methylation status is available.  

\begin{table}[!t]
\small
\caption{Qualitative (color) comparison of tissue appearances on different MRI modalities.}
\label{tab1}
\centering
\begin{tabular}{lllll}
\hline
\textbf{Tissue} & \textbf{FLAIR} & \textbf{T1w} & \textbf{T1wCE} & \textbf{T2w} \\
\hline 
\textbf{CSF} & Dark & Dark & Dark & Bright\\

\textbf{White Matter} & Dark Gray & Light & Light Gray & Dark Gray\\

\textbf{Cortext} & Light Gray & Gray & Light Gray & Light Gray\\

\textbf{Inflammation} & Bright & Dark & Bright & Bright\\

\textbf{Fat} & Light & Bright & Light Gray & Light\\
\hline
\end{tabular}
\end{table}

\begin{figure*}
  \centering
    \begin{subfigure}{0.7\textwidth} % <----
     \includegraphics[width=\textwidth]{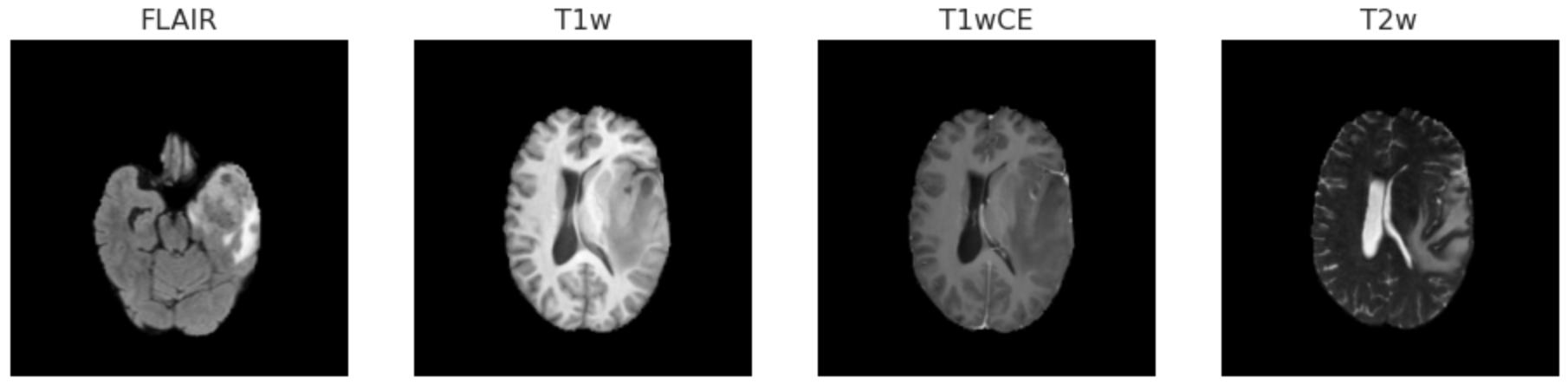}
     \caption{Patient with MRIs in the same orientation plane}
     \label{fig1a}
    \end{subfigure}

    \begin{subfigure}{0.7\textwidth}  % <----
     \includegraphics[width=\textwidth]{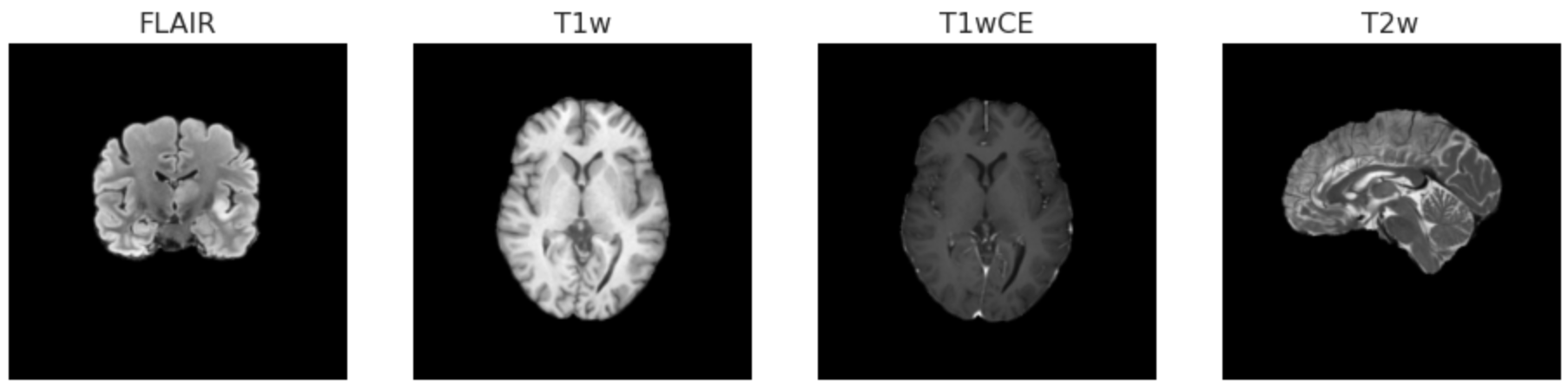}
     \caption{Patient with MRIs in different orientation planes}
     \label{fig1b}
    \end{subfigure}

\caption{Examples of patients having MRI scans in the same and different orientations. We show in this example for one patient (a), the different types of MRI scans (i.e. FLAIR, T1w, T1wCE, and T2w) are all saved in the same axial view, but for another patient (b), the different types of MRI scans are saved in different orientations (i.e. FLAIR is in the coronal view, T1w and T1wCE in the axial view, and T2w in the sagittal view).}
\label{fig:scans_ex_orie}
\end{figure*}

\begin{figure*}[!t]
\centering
\begin{tabular}{c|c}
    {\includegraphics[width=0.88\columnwidth]{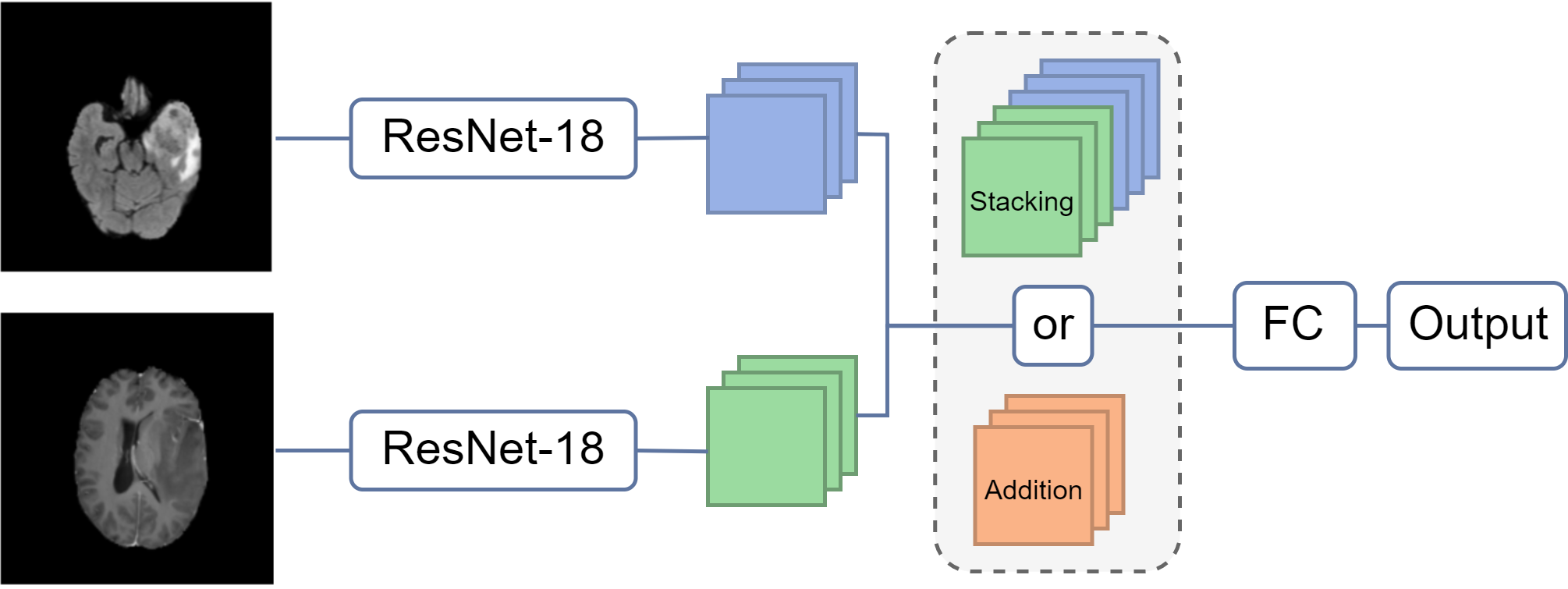}}&

    {\includegraphics[width=1.12\columnwidth]{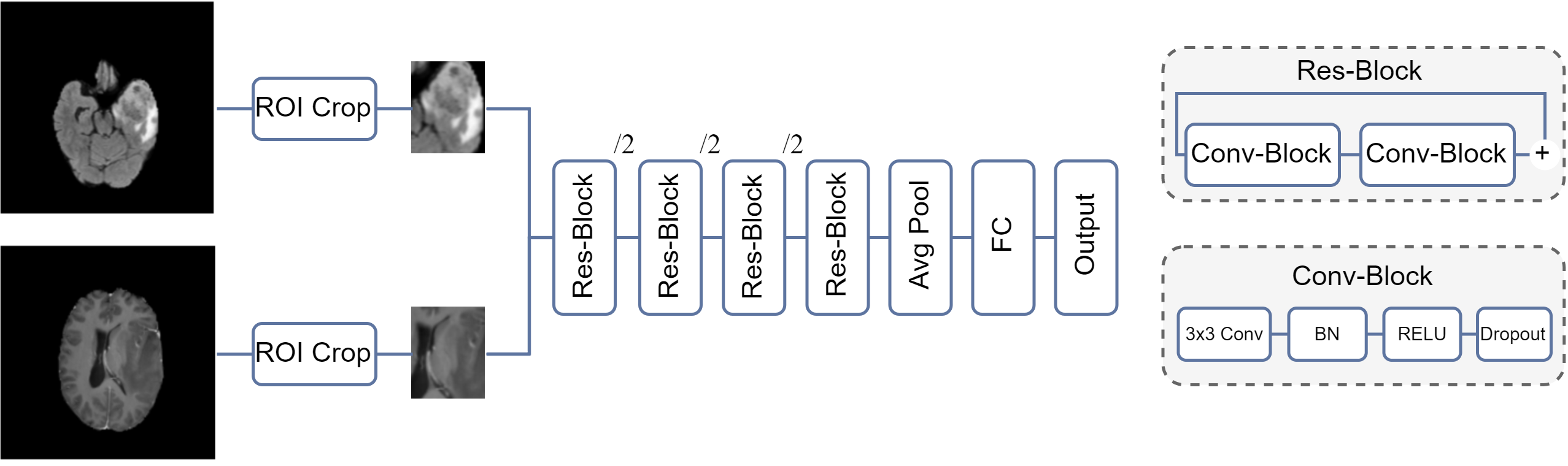}}
    \\
    (a)&(b)
\end{tabular}
\caption{(a) A custom ResNet architecture for merging FLAIR and T1wCE modalities. (b) A custom ResNet architecture for ROI-based classification, as described in \cite{chang2018deep}. Each residual block consists of two convolutional blocks with a dropout layer added. After each stage, feature maps are downsampled using convolution with stride 2 (indicated by /2). Before Fully Connected (FC) layer, 4 x 4 average pooling is utilized.}
\label{fig:custom_experiments}
\end{figure*}

The radiogenomic classification task dataset contains multi-parametric MRI (mpMRI) scans for 585 patients with glioblastoma. Patients fall into two categories based on their MGMT status: methylated MGMT (1) and unmethylated MGMT (0). The dataset is evenly distributed across the two classes, with 307 methylation cases and 278 unmethylated cases, as shown in Figure \ref{fig:MRI-dist}(a). As a result of the balanced nature of the dataset, the learning algorithm is not expected to show skewing behavior. In addition, metadata such as modality, orientation, and MRI machine-specific information are included in the Digital Imaging and Communications in Medicine (DICOM) header of the images. Images have variable slice thicknesses ranging from 0.43 to 6 mm. Four imaging modalities are used to capture each patient's images: T1-weighted pre-contrast (T1), T1-weighted post-contrast (T1wCE (Gadolinium)), T2-weighted (T2), and T2 Fluid Attenuated Inversion Recovery (T2-FLAIR). Each of these MRI modalities highlights a different aspect of the target area which, when used together, can assist in better localizing the tumor region than can one modality alone. Table \ref{tab1} depicts the differences between the MRI modalities. 

The number of slices for each patient differs across the modalities mentioned above. Patients with the IDs [00109], [00123], and [00709] are excluded from the dataset because the quality of their images is unacceptable and we do not want them to affect our analysis. Furthermore, MRI scans come in three orientation planes, including coronal, axial, and sagittal. Figures \ref{fig:MRI-dist}(b) and \ref{fig:MRI-dist}(c) display the distribution of scans among various MRI modalities and orientations, respectively. The different scan modalities for some patients were saved in the same orientation plane, but not for all patients. Figure \ref{fig:scans_ex_orie} shows four random slices for two patients displaying the four modalities and the existence of the same and different orientations for different patients within the dataset.

In this work, we used data from both tasks separately: 1) Segmentation Task and 2) Classification Task. We extracted the MRI scans of the patients present in the classification task from the dataset of the segmentation task through their patient ID. The classification dataset provides MRI scans in its raw form; however, the segmentation task dataset is standardized by applying the following pre-processing steps: 

\begin{enumerate}
\item DICOM files are converted to NIFTI file format.
\item Re-orientation to have a common orientation (i.e., RAI).
\item Co-registration to the same anatomical template (SRI24).
\item Skull-stripping.
\end{enumerate}

The pre-processing pipeline used for the Task 1 dataset is publicly available through the Cancer Imaging Phenomics Toolkit (CaPTk) \citep{davatzikos2018cancer}. The classification dataset is preprocessed by resampling the scans to the same axial plane for consistency and, following the method used by the winning solution of the Kaggle competition, extracting the same number of slices for each patient. . We then use the segmentation and the classification dataset (separately) to train the different deep-learning models throughout the study. The main reason behind using Task 1 (Segmentation) dataset is that it is already pre-processed with no discrepancies in registration, orientation, and the number of slices across patients. Thus, by comparing the performance of the deep-learning models on both datasets, we can eliminate the possibility of any flaws in our pre-processing implementation for the Task 2 dataset.

\section{Experimental Methods}
\label{experiments}
We conducted extensive experiments involving two main families of deep network architectures: convolutional neural networks (CNNs) and transformers. 

\subsection{Convolutional neural networks}
CNNs have seen many developments and have long been the go-to for computer vision tasks. Their ability to tackle scale invariance and extract generalizable features are primarily credited to their inductive biases that help model local visual structures. Their successes in medical classification have been reported in many works, including \cite{skin_lesion_classif, breast_tumor_classif, cnn_ssl_classif, atwany2022drgen, ridzuan2022self}, and others.

We begin our experiments using ResNet and its variations. \cite{resnet} introduced a bypass skip connection that enables cross-layer connectivities to mitigate the vanishing gradient problem. We perform our experiments using ResNet-10, -18, and -34. Despite its small network depth, ResNet-10, in particular, has been recorded as the winning solution for BraTS2021 \citep{brats2021}.

Building upon ResNet, \cite{densenet} extends the application of skip connections to the layer level, where each layer is concatenated to the subsequent ones densely, thus giving a name to the DenseNet architecture. This allows for a reduction in the number of parameters through feature reuse and feature propagation. We experiment with DenseNet-121, -169, and -201.

To improve computational performance and accuracy, \cite{efficientnet} used a compound scaling coefficient to uniformly scale and balance three dimensions of the model network: architecture depth, architecture width, and image resolution. They introduced the EfficientNet architecture, from which we employ EfficientNet-b0, -b1, and -b3.

\subsection{Transformers}
Despite their efficiency, the performance of CNNs is arguably impaired by their inability to capture long-range spatial relationships. \cite{attention_all_you_need} overcame this by introducing the self-attention mechanism in transformers that allow the model to learn long-range dependencies for better representation learning. We investigate the use of Vision Transformer (ViT) \citep{vit} and Shifted window (Swin) \citep{swin} transformer for brain tumor classification.
In ViTs \citep{vit}, an image is partitioned into a sequence of 16 $\times$ 16 patches that is fed into a transformer encoder. \cite{swin} improved upon this architecture by introducing the inductive bias of CNNs using a hierarchical structure and a locally shifted window within which self-attention is computed.

\subsection{Pretraining}
Given the scarcity of labeled data, we also examine the effect of self-supervision (SSL) on the classification of brain tumors from MRIs. \cite{cnn_vs_transformer} showed that an ImageNet \citep{imagenet} pre-training could favorably boost the performance of ViTs in medical images, rivaling that of CNNs. We thus investigate using a contrastive-based self-supervised learning method, SimCLR \citep{simclr}, to pre-train a model with a proxy task before fine-tuning it on the downstream classification.

Experiments are carried out using MONAI and Pytorch. All classification models are trained using AdamW optimizer on a single NVIDIA RTX A6000 GPU.

\begin{table}[!t]
\small
\caption{\label{ssltable}Comparison of results on pretraining different models using ImageNet pretrained weights and contrastive learning. FLAIR is used for ImageNet weight initialization. For SimCLR, we used all four modalities so that the model could learn more during the pretraining stage and generalize well during finetuning. Validation AUC represents the AUC values on the held-out testing set comprising 20\% of the data.}
\centering
\begin{tabular}{ccccc} 
\hline
Model & Modality & Pretraining & Val. AUC  \\
\hline
ResNet-50 & FLAIR & ImageNet & 0.58          \\
DenseNet-121 & FLAIR & ImageNet & 0.58          \\
ResNet-18 & All & SimCLR & 0.60                \\
\hline
\end{tabular}
\end{table}

\begin{table*}[!t]
\small
\caption{\label{tab2}Comparison of five-fold cross-validation AUCs on different MRI modalities using the BRaTS2021 Task 2 dataset. The last column shows the mean and variance of the five folds. The largest AUC by fold and the largest mean AUC are bolded. The mean AUC saturates around 0.62, suggesting that the models are unable to differentiate between the methylated statuses of the tumor patients.}
\centering
\begin{tabular}{p{2.5cm}p{1.5cm}cccccc} 
\hline
Model & Modality & Fold 1 & Fold 2 & Fold 3 & Fold 4 & Fold 5 & Mean $\pm$ Variance AUC\\
\hline
ResNet-10 & T1wCE & \textbf{0.603}  & 0.557  & 0.539  & 0.595  & 0.547  & 0.5682 $\pm$ 0.0008 \\
ResNet-18 & T1wCE & 0.575  & 0.596  & 0.562  & 0.647  & 0.595  & 0.5950 $\pm$ 0.0010 \\
ResNet-34 & T1wCE & 0.594  & 0.593  & 0.550   & 0.670   & 0.585  & 0.5984 $\pm$ 0.0019 \\
ResNet-10 & FLAIR & 0.593  & 0.616  & 0.565  & 0.678  & 0.570   & 0.6044 $\pm$ 0.0021 \\
ResNet-18 & FLAIR & \textbf{0.603}  & 0.599  & \textbf{0.630}   & 0.656  & \textbf{0.596}  & 0.6168 $\pm$ 0.0007 \\
ResNet-34 & FLAIR & 0.577 & \textbf{0.623} & 0.629 & \textbf{0.680} & 0.580 & \textbf{0.6178} $\pm$ 0.0018
\\
\hline
\end{tabular}
\end{table*}

\begin{table*}[!t]
\small
\caption{\label{tab3}Comparison of five-fold cross-validation AUCs on different MRI modalities using the BRaTS2021 Task 1 dataset. The last column shows the mean and variance of the five folds. The largest AUC by fold and the largest mean AUC are bolded. The mean AUC saturates around 0.63, suggesting that the models are unable to differentiate between the methylated statuses of the tumor patients.}
\centering
\begin{tabular}{p{2.5cm}p{1.5cm}cccccc} 
\hline
Model & Modality & Fold 1 & Fold 2 & Fold 3 & Fold 4 & Fold 5 & Mean $\pm$ Variance AUC\\
\hline
ResNet-10 & T1wCE & 0.5568 & 0.5646 & 0.5744 & 0.5892 & 0.6071 & 0.5784 $\pm$ 0.0004\\

ResNet-18 & T1wCE & 0.5845 & 0.5610 & 0.6124 & 0.5713 & 0.5858 & 0.5830 $\pm$ 0.0004\\

ResNet-34 & T1wCE & 0.5970 & 0.5810 & \textbf{0.6476} & 0.5886 & 0.6688 & 0.6166 $\pm$ 0.0015\\

DenseNet-121 & T1wCE & 0.6072 & 0.6116 & 0.6067 & 0.6228 & \textbf{0.6901} & 0.6277 $\pm$ 0.0013\\

DenseNet-169 & T1wCE & 0.6155 & 0.5813 & 0.5568 & 0.6857 & 0.5944 & 0.6067 $\pm$ 0.0024\\

DenseNet-201 & T1wCE & 0.5714 & 0.5518 & 0.5977 & 0.6441 & 0.5954 & 0.5913 $\pm$ 0.0016\\

EfficientNet-b0 & T1wCE & \textbf{0.6581} & 0.6192 & 0.6127 & 0.5997 & 0.6090 & 0.6197 $\pm$ 0.0005\\

EfficientNet-b1 & T1wCE & 0.6003 & 0.5402 & 0.5932 & 0.6059 & 0.5457 & 0.5771 $\pm$ 0.0010\\

EfficientNet-b3 & T1wCE & 0.6092 & 0.5432 & 0.6003 & 0.6533 & 0.5698 & 0.5952 $\pm$ 0.0017\\

\\

ResNet-10 & FLAIR & 0.5967 & 0.5634 & 0.6137 & 0.4000 & 0.5855 & 0.5519 $\pm$ 0.0075\\

ResNet-18 & FLAIR & 0.5967 & 0.5402 & 0.6102 & 0.6154 & 0.5818 & 0.5889 $\pm$ 0.0009\\

ResNet-34 & FLAIR & 0.5908 & 0.5693 & 0.6188 & 0.5196 & 0.5818 & 0.5761 $\pm$ 0.0013\\

DenseNet-121 & FLAIR & 0.5773 & 0.5845 & 0.5964 & 0.6136 & 0.5772 & 0.5898 $\pm$ 0.0002\\

DenseNet-169 & FLAIR & 0.6030 & 0.6110 & 0.5539 & 0.6632 & 0.5793 & 0.6021 $\pm$ 0.0017\\

DenseNet-201 & FLAIR & 0.6155 & 0.6071 & 0.5273 & 0.6379 & 0.5645 & 0.5970 $\pm$ 0.0023\\

EfficientNet-b0 & FLAIR & 0.5720 & 0.6006 & 0.5472 & 0.5818 & 0.5478 & 0.5699 $\pm$ 0.0005\\

EfficientNet-b1 & FLAIR & 0.6128 & 0.5810 & 0.6300 & 0.6555 & 0.5546 & 0.6068 $\pm$ 0.0016\\

EfficientNet-b3 & FLAIR & 0.5893 & 0.5482 & 0.5568 & 0.5918 & 0.5870 & 0.5746 $\pm$ 0.0004\\

\\

DenseNet-121 & T2 & 0.5627 & 0.5914 & 0.5977 & \textbf{0.6934} & 0.6046 & 0.6100 $\pm$ 0.0024\\

DenseNet-169 & T2 & 0.5738 & 0.6193 & 0.5737 & 0.6663 & 0.6083 & 0.6083 $\pm$ 0.0015\\

DenseNet-201 & T2 & 0.5955 & 0.6128 & 0.6469 & 0.6108 & 0.5880 & 0.6108 $\pm$ 0.0005\\

EfficientNet-b0 & T2 & 0.6203 & 0.6068 & 0.6242 & 0.6726 & 0.5975 & 0.6243 $\pm$ 0.0008\\

EfficientNet-b1 & T2 & 0.5937 & 0.6098 & 0.6047 & 0.6724 & 0.6269 & 0.6215 $\pm$ 0.0010\\

EfficientNet-b3 & T2 & 0.6167 & 0.5923 & 0.5744 & 0.6700 & 0.6052 & 0.6117 $\pm$ 0.0013\\

\\

ResNet-10 & All & 0.5204 & 0.5300 & 0.5321 & 0.5263 & 0.5731 & 0.5364 $\pm$ 0.0004\\

ResNet-18 & All & 0.6387 & 0.5360 & 0.5216 & 0.5692 & 0.5824 & 0.5696 $\pm$ 0.0021\\

ResNet-34 & All & 0.5923 & 0.5711 & 0.6421 & 0.6176 & 0.6361 & 0.6118 $\pm$ 0.0009\\

DenseNet-121 & All & 0.5702 & 0.5735 & 0.5126 & 0.6253 & 0.6355 & 0.5834 $\pm$ 0.0024\\

DenseNet-169 & All & 0.5493 & 0.5717 & 0.6105 & 0.6256 & 0.6194 & 0.5953 $\pm$ 0.0011\\

DenseNet-201 & All & 0.5848 & 0.6131 & 0.5539 & 0.5861 & 0.6293 & 0.5934 $\pm$ 0.0008\\

EfficientNet-b0 & All & 0.6292 & 0.5759 & 0.5705 & 0.6153 & 0.6046 & 0.5991 $\pm$ 0.0006\\

EfficientNet-b1 & All & 0.6095 & \textbf{0.6634} & 0.5999 & 0.6687 & 0.6130 & \textbf{0.6309} $\pm$ 0.0011\\

EfficientNet-b3 & All & 0.5878 & 0.6021 & 0.5803 & 0.6163 & 0.5750 & 0.5923 $\pm$ 0.0003\\

ViT & All & 0.5583 & 0.5810 & 0.5000 & 0.5100 & 0.5963 & 0.5491 $\pm$ 0.0018\\

SwinViT & All & 0.5943 & 0.6144 & 0.5863 & 0.5255 & 0.5285 & 0.5698 $\pm$ 0.0016\\
\hline
\end{tabular}
\end{table*}

\section{Results} 
Results for all the experiments are reported in Tables~\ref{tab2} and~\ref{tab3}. Table~\ref{tab2} lists the results using the dataset provided for Task 2 (classification). Note that this dataset comes in a raw format with no pre-processing techniques applied. Table~\ref{tab3} shows the results for Task 1 dataset (segmentation) with publicly available pre-processing steps as mentioned above. All the experiments are carried out using \textit{k}-fold cross-validation (where \textit{k}=5) and using a standalone modality or a combination of modalities as listed in the tables. This approach is preferred due to the nature of T1wCE and FLAIR modalities. These modalities are utilized in most experiments since the tumor's contrast is high and appears bright. It is, therefore, easier for the models to distinguish the tumor region, which is hypothesized to encode information regarding the MGMT promoter status.

\subsection{Results on Data from Task 2}
The Task 2 dataset is in a raw DICOM format, and thus in-house preprocessing techniques are applied before training. Table~\ref{tab2} shows the AUC values of 5-fold cross-validation on Task 2 (classification) dataset along with their mean and variance. As mentioned, the ResNet family (ResNet-10, -18, and -34) is analyzed in this set of experiments. With a slight discrepancy between the modalities and models, all AUC values ranged between 0.5682 and 0.6178. The variance between the results is also similar, indicating that the models cannot properly perform the classification task.

\subsection{Results on Data from Task 1 }
Assuming that the first dataset's pre-processing steps might be flawed, we use the Task 1 dataset dedicated to segmentation. As mentioned, pre-processing techniques are available online for this task, and as such, there are no inconsistencies and inaccuracies in the standardization steps. For this set of experiments, T1wCE, FLAIR, T2, and all modalities combined are utilized to mimic the clinicians' routine where they might favor one modality over another or combine all modalities together to make their decisions. DL families explored for this part were ResNet, DenseNet, EfficientNet, and ViTs. The reasoning behind the choices of networks is provided in Section~\ref{experiments}. ResNet-10, -18, -34 models are tested with T1wCE, FLAIR, and all modalities combined datasets; DenseNet-121, -169, -201, and EfficientNet-b0, -b1, -b3 models are used for all the modalities individually and combined. Finally, a ViT model is also utilized for the combined dataset. As seen in the table, the AUC scores are around 0.59 regardless of the depth or complexity of the models. A similar pattern is also seen consistent with the ViT model, with a mean AUC of 0.549 when all the modalities are combined for training. Results of Swin Transformer experiments reached the same performance, with a mean AUC of 0.5698. The mean AUC scores for all experiments lie in the range of 0.536-0.631, indicating that no model can satisfactorily perform the classification task using the MR imaging data, regardless of how simple or complex the model is. Such AUC values are inauspicious, especially when actual clinical applications are concerned.

\begin{table*}[!t]
\small
\caption{\label{tweakstable}Comparison of different configurations and their results for custom model training. Ensemble models refer to two-path ResNet, each of which is responsible for FLAIR and T1wCE modalities individually for feature extraction as depicted in Figure~\ref{fig:custom_experiments}(a). ROI-based custom model represents the custom ResNet from~\citep{chang2018deep} as illustrated in Figure~\ref{fig:custom_experiments}(b) that relies on the region-of-interest (i.e. tumor region) for prediction. Fusion technique shows how the modalities are treated during the training process. Image resolution correspond to the resized image sizes for the training. Validation AUC are the results on the held-out testing set comprising 20\% of the data.}
\centering
\begin{tabular}{ccccc} 
\hline
Model & Modality & Fusion Technique & Image Res. & Val. AUC\\
\hline
Ensemble ResNet-18 & FLAIR \& T1wCE & Addition & 128  & 0.49\\
Ensemble ResNet-18 & FLAIR \& T1wCE & Stacking & 128  & 0.63\\
ROI-based Custom ResNet & T1wCE & Single-channel & 128 & 0.55\\
ROI-based Custom ResNet & FLAIR \& T1wCE & Two-channel & 32  & 0.53\\
\hline
\end{tabular}
\end{table*}

\subsection{Results on Custom Models}
We conduct several additional experiments with different tweaks. Their summary is provided in Table~\ref{tweakstable}. First, we use an ensemble of two ResNet-18 models without FC layers, as shown in Figure~\ref{fig:custom_experiments}(a). Each model is utilized to extract features from FLAIR and T1wCE in separate parallel paths and concatenated or added before feeding into the FC layers for MGMT classification. When we apply addition to the extracted features, the customized network achieve an AUC of 0.49; when stacking is used, it reaches 0.63 AUC. Another set of experiments carried out is based on the region-of-interest (ROI), as depicted in Figure~\ref{fig:custom_experiments}(b). Segmentation masks from the Task 1 dataset are used to extract only the tumor region (as ROI) in a cropped slice-by-slice fashion with the size of $32\times32$ as in~\cite{chang2018deep} and $128\times128$ for additional exploration. We use T1wCE with the size of $128\times128$ as an input, reaching 0.55 AUC. With the size of $32\times32$, FLAIR and T1wCE are both used as input to achieve an AUC of 0.53. In all the experiments with different tweaks, the results follow the same trend with an AUC of around 0.5-0.6 despite the variations introduced in the models, input data, customization, and training methods.

\subsection{Results on Pretraining}
Because pretraining is typically praised for improving model learning, we investigate the performance of CNN models using pretrained weights. Specifically, we perform contrastive learning-based SSL pretraining as summarized in Table~\ref{ssltable}. ResNet-50 and DenseNet-121, initialized with ImageNet weights, show AUC scores of 0.58 and 0.58 on the Task 2 dataset. For contrastive learning, we use SimCLR~\citep{simclr} to maximize the model's understanding of the variations in the MRI scans. For this experiment, we utilize the BRaTS2020 dataset to perform the pretraining of ResNet-18 and finetuned the model on Task 2 dataset. The reason for this approach is to avoid any data leakage and also use two different yet semantically similar datasets for pretraining and finetuning, respectively. It thus ensures that the model learns discriminative features during pretraining while validating rigorously during finetuning. SSL-based initialization results are not much different, with an AUC of 0.6. 

These findings are contrary to the optimistic results reported in several previous papers. To understand such performance discrepancies, we investigate the models, modalities, loss performances, and overall interpretability of the models.

\begin{figure*}[!t]
  \centering
     \includegraphics[width=0.75\textwidth]{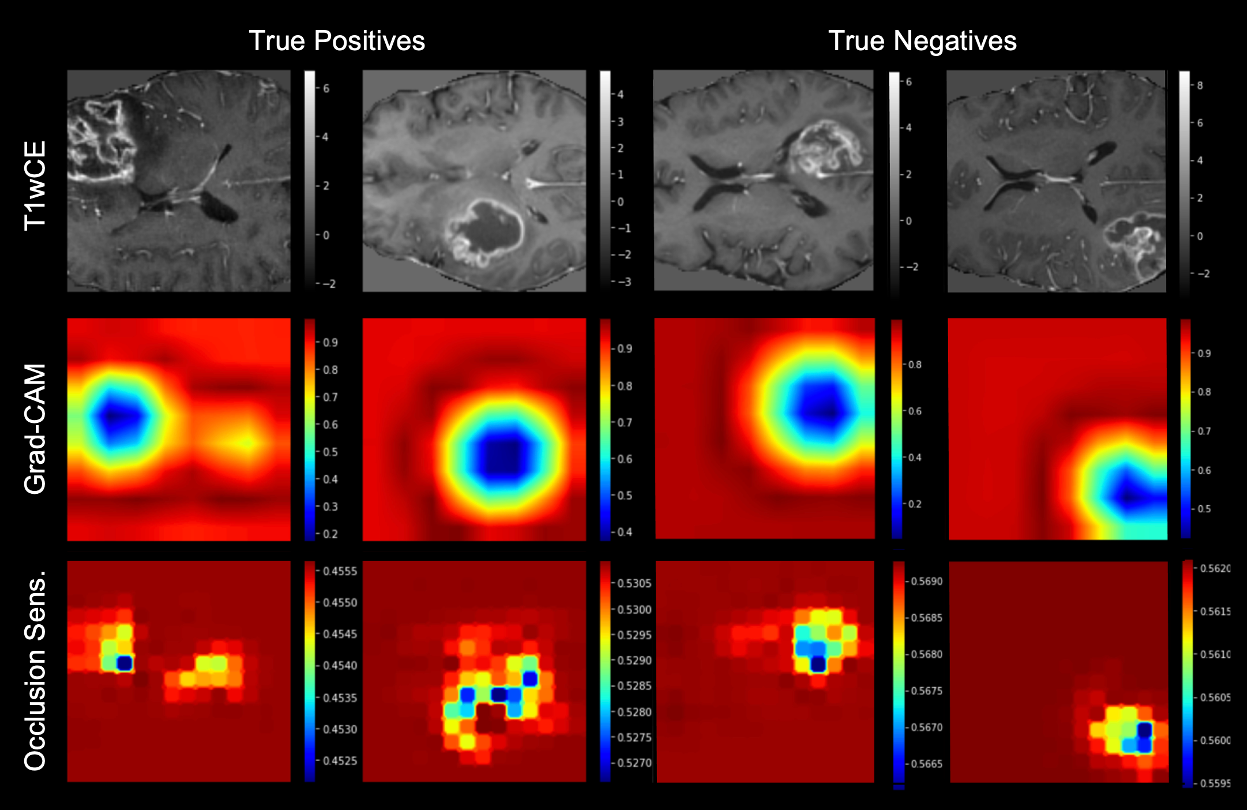}
     \caption{Saliency maps of the successful predictions of ResNet-10. Top to bottom: Original T1wCE central slices from the axial plane; Grad-CAM heatmaps from the activation map of the last convolutional layer of the model; occlusion sensitivity maps using a mask size and stride of 12. The heatmap importance increases from red to blue. Columns 1-2: Correct predictions of methylated tumor patients. Columns 3-4: Correct predictions of non-methylated tumor patients.}
\label{fig:brats_gradcam_occ_sensitivity_true_pos_true_neg}
\end{figure*}

\begin{figure*}[!t]
  \centering
     \includegraphics[width=0.75\textwidth]{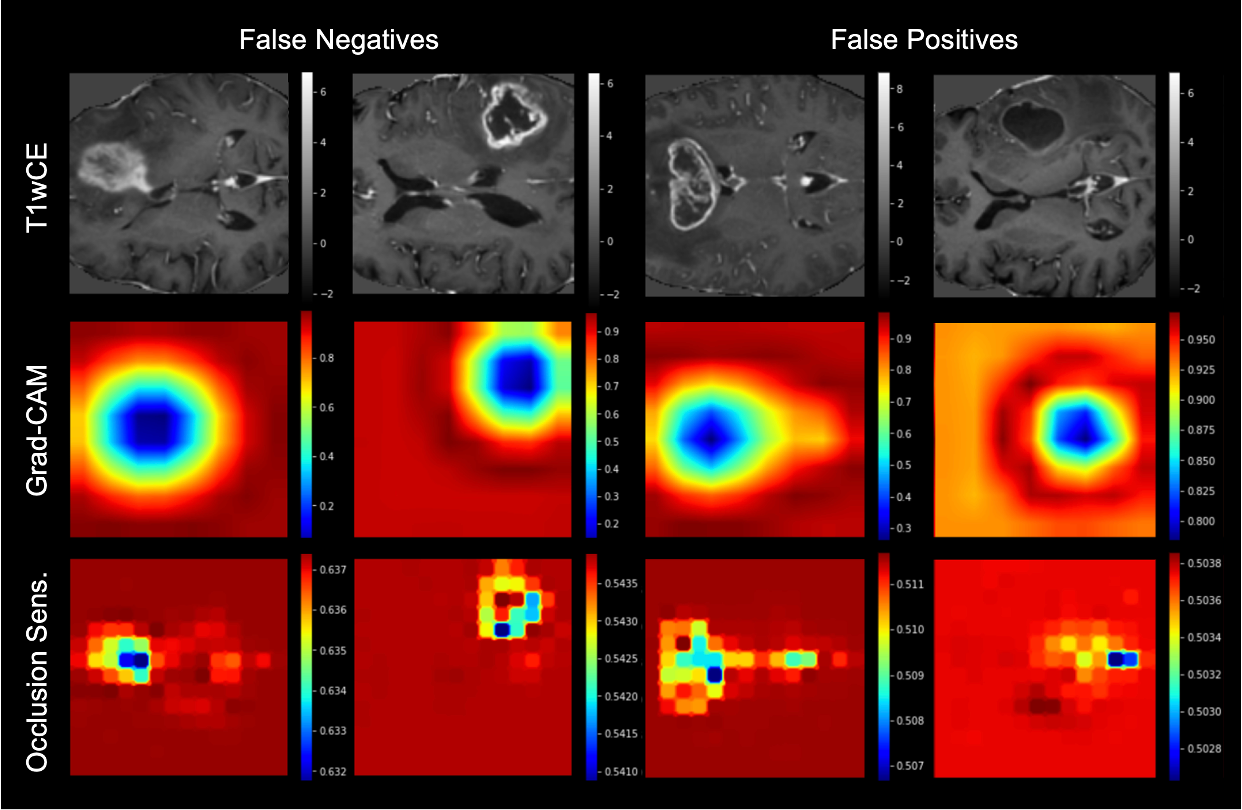}
     \caption{Saliency maps of the failed predictions of ResNet-10. Top to bottom: Original T1wCE central slices from the axial plane; Grad-CAM heatmaps from the activation map of the last convolutional layer of the model; occlusion sensitivity maps using a mask size and stride of 12. The heatmap importance increases from red to blue.  Columns 1-2: Incorrect predictions of methylated tumor patients. Columns 3-4: Incorrect predictions of non-methylated tumor patients.}
\label{fig:brats_gradcam_occ_sensitivity_false_neg_false_pos}
\end{figure*}

\section{Discussion} 
\label{section:discussion}
In this work, we have examined the capability of state-of-the-art deep learning models and training methods to predict the MGMT promoter status reliably. We find that regardless of the model and the training methodologies used, the prediction power of the trained models is very low. To better understand the behavior of the trained models, we thoroughly evaluate their performance. We attempt to tap into their inner dynamics using the interpretability methods of Grad-CAM \citep{gradcam}, occlusion sensitivity \citep{occ_sens}, t-SNE \citep{van2008visualizing}, PCA \citep{pearson1901_pca}, probability distribution, and loss landscape \citep{li2018visualizing}. 
%consider introducing the lung dataset here and mention you will eventually compare its performance with the BRATs dataset. 

\subsection{Grad-CAM and Occlusion Sensitivity}
Grad-CAM \citep{gradcam}, or gradient-weighted class activation map, makes use of the gradients of the target output with respect to the final convolutional layer to produce a coarse localization map that highlights the regions of the image responsible for predicting the target. Figures \ref{fig:brats_gradcam_occ_sensitivity_true_pos_true_neg} and \ref{fig:brats_gradcam_occ_sensitivity_false_neg_false_pos} (middle row) display the Grad-CAM outputs of ResNet-10 for the correct and incorrect predictions of the model, respectively. Interestingly, in many cases, the model appears to be localizing the tumor regions (despite its wrong final prediction). However, occasionally (e.g., Figure \ref{fig:brats_gradcam_occ_sensitivity_false_neg_false_pos}, last column), it misidentifies the region of interest. The occlusion sensitivity analysis further supports this observation.

In occlusion sensitivity \citep{occ_sens}, a small perturbation is created on the input image using an occluded mask. As the mask moves across the image, the probability of the inferred class changes accordingly. A sharp fall in probability score indicates that the occluded region is vital in inferring the given class. The occlusion sensitivity maps (Figures \ref{fig:brats_gradcam_occ_sensitivity_true_pos_true_neg} and \ref{fig:brats_gradcam_occ_sensitivity_false_neg_false_pos}, bottom row) show that the model is detecting abnormalities in the brain, consistent with the Grad-CAM findings. However, it is observed that despite the tumor's successful localization, the model could not detect features within the tumor that could possibly relate to discriminating the MGMT promoter methylation status.

\begin{figure*}[t!]
    \centering
    \includegraphics[width=0.8\textwidth]{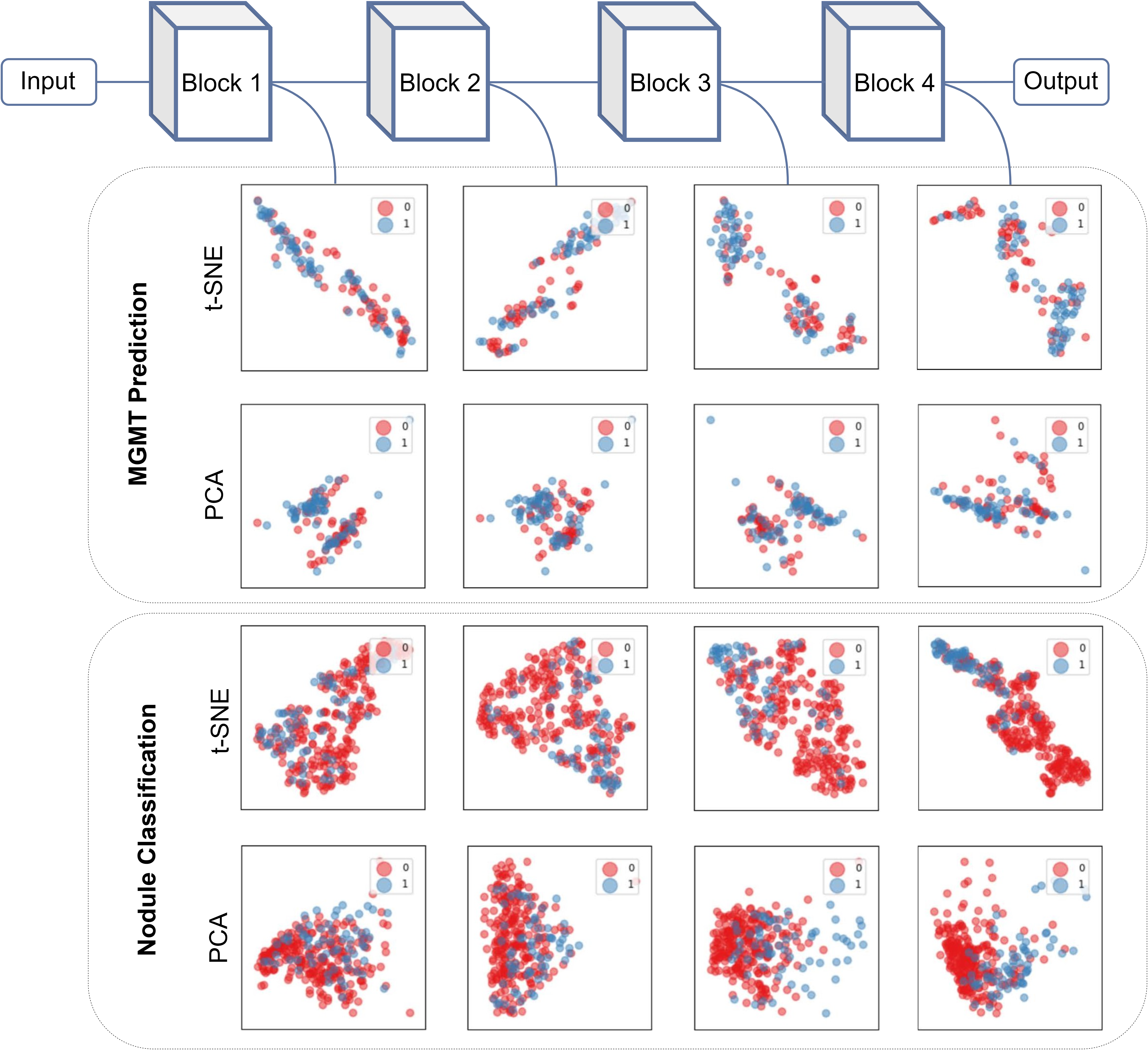}
    \caption{t-SNE and PCA dimensionality reduction visualization of the feature maps progressing through the different ResNet-10 layers using the BRaTS2021 (top) and NoduleMNIST3D (bottom) medical datasets. 0 is non-methylated and 1 is methylated for BRaTS2021; 0 is benign and 1 is malignant lung nodules for NoduleMNIST3D. Towards the final layers, a clearer separation between the binary classes becomes apparent on NoduleMNIST3D but not on BRaTS2021.}
    \label{fig:tsne}
\end{figure*}

\begin{table*}[!th]
\small
\caption{\label{AUC_nodule}Five-fold cross-validation AUCs on NoduleMNIST3D. The last column shows the mean and variance of the five folds. This experiment is performed with ResNet-10 to compare the visualization of feature maps of the model on BRaTS and NoduleMNIST3D datasets, as described in Section~\ref{featuremaps}. The same model for the same task with two datasets perform with a high discrepancy as is visually validated in Figure~\ref{fig:tsne}.}
\centering
\begin{tabular}{p{2.5cm}cccccc} 
\hline
Model  & Fold 1 & Fold 2 & Fold 3 & Fold 4 & Fold 5 & Mean $\pm$ Variance AUC\\
\hline
ResNet-10  & 0.865 & 0.863 & 0.848 & 0.853 & 0.852 & 0.856 $\pm$ 0.007\\
\hline
\end{tabular}
\end{table*}

\subsection{Feature Maps of CNN}
\label{featuremaps}
To validate the above argument, feature maps are extracted at each layer of the trained ResNet-10 model based on the validation set. However, due to the features' high dimensionality, it is impossible to visualize them directly. Therefore, after applying average pooling on the feature maps, we use t-SNE and PCA to reduce the dimensionality of the feature maps and plot them at every layer in 2D space. In Figure \ref{fig:tsne}, it can be observed that towards the final layers, the two classes are still entangled, suggesting that the model cannot find features that can differentiate between the two classes. To compare it with another model trained for the binary classification of a medical disease using a 3D dataset, we train a ResNet-10 model using the NoduleMNIST3D dataset \citep{medMNIST, armato2011lung} to classify the malignancy of nodules in lungs. The classification performance of the model on this dataset can be found in Table \ref{AUC_nodule}. It can be observed from the t-SNE and PCA plots in Figure \ref{fig:tsne} that the trained ResNet-10 model is able to extract separable features from the region of interest and consequently is able to separate the different classes apart in the final layers.

\subsection{Probability Distribution of the Predictions}
\begin{figure*}[!t]
\centering
\begin{tabular}{rccc}
    {\includegraphics[width=0.6\columnwidth]{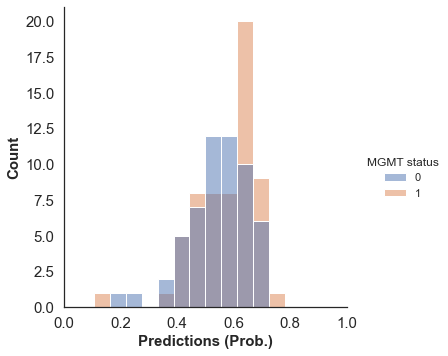}}&
    {\includegraphics[width=0.6\columnwidth]{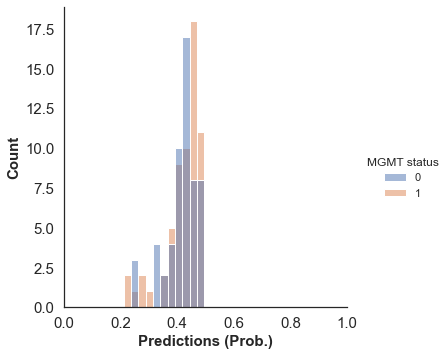}}&
    {\includegraphics[width=0.6\columnwidth]{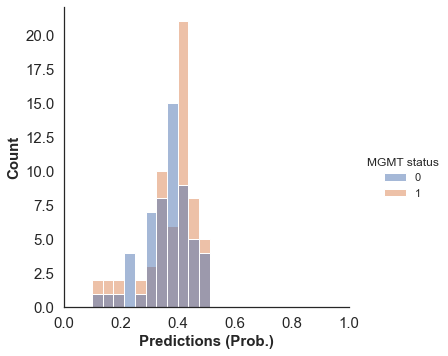}}
    \\
    &(a)& \\
    {\includegraphics[width=0.55\columnwidth]{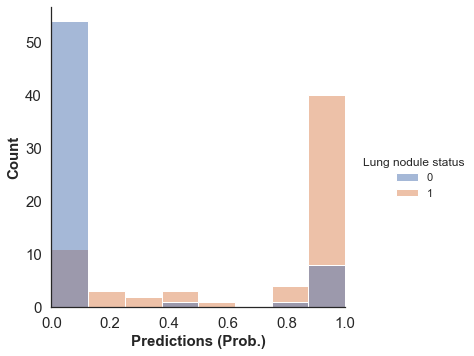}}&
    {\includegraphics[width=0.55\columnwidth]{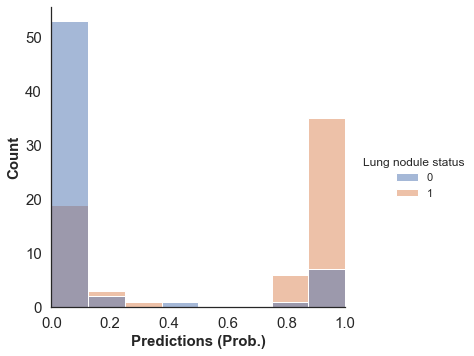}}&
    {\includegraphics[width=0.55\columnwidth]{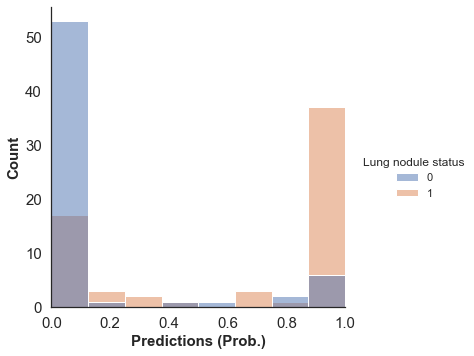}}
    \\
    &(b)&
\end{tabular}

\caption{Histogram of the predicted classification probabilities conditioned on true labels. (a) shows the predicted probabilities conditioned on the true labels for three random experiments on BRaTS2021 dataset. Orange bars (1) represent methylated and blue bars (0) correspond to unmethylated MGMT promoter, while gray is the overlap between the two. (b) shows the predicted probabilities conditioned on the true labels for three random experiments on the NoduleMNIST3D dataset. Orange bars (1) represent malignant, blue bars (0) correspond to benign lung nodules, while gray is the overlap between the two.}
\label{fig:histogram} \label{histogram}
\end{figure*}

Furthermore, we examine the probability distribution of the ResNet-10 prediction for both the BRaTS and the NoduleMNIST3D datasets. For binary classification, a well-trained model is expected to exhibit a bimodal distribution of the prediction probabilities, showing some confidence in the predicted class (probabilities being close to 0 or 1). As can be observed from Figure \ref{histogram}a, the model exhibits a unimodal distribution with a mean localized between 0.4-0.6 suggesting uncertainty in the final prediction. In comparison, the same model trained on NoduleMNIST3D exhibits a bimodal distribution implying higher confidence in separating the two classes (Figure \ref{histogram}b).

\subsection{Loss Landscape}
\begin{figure*}[!htbp]
\captionsetup[subfigure]{justification=centering}
\centering
\begin{minipage}{0.24\textwidth}
\begin{subfigure}{\textwidth}
    \includegraphics[width=\textwidth]{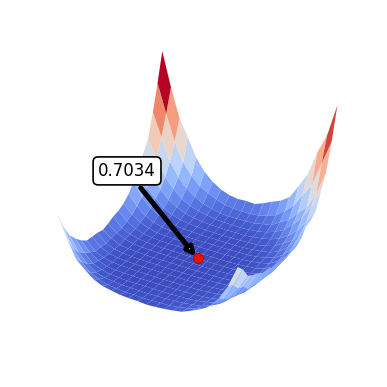}
    \subcaption{\textbf{T1wCE}}
\end{subfigure}
\end{minipage}
\begin{minipage}{0.24\textwidth}
\begin{subfigure}{\textwidth}
    \includegraphics[width=\textwidth]{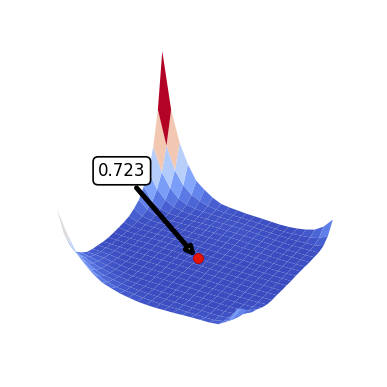}
    \subcaption{\textbf{FLAIR}}
\end{subfigure}
\end{minipage}
\begin{minipage}{0.24\textwidth}
\begin{subfigure}{\textwidth}
    \includegraphics[width=\textwidth]{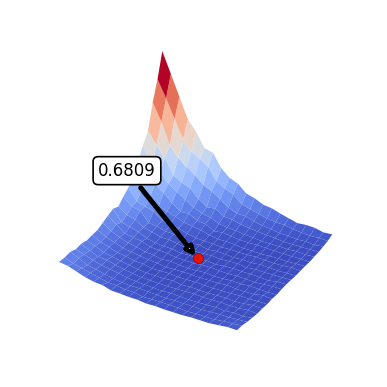}
    \subcaption{\textbf{Multi-Modal}}
\end{subfigure}
\end{minipage}
\caption{Binary cross-entropy loss landscapes of ResNet-34 model on the BRaTS2021 dataset.}
\label{data_sample}
\end{figure*}

\begin{figure*}[!htbp]
\captionsetup[subfigure]{justification=centering}
\centering
\begin{minipage}{0.24\textwidth}
\begin{subfigure}{\textwidth}
    \includegraphics[width=\textwidth]{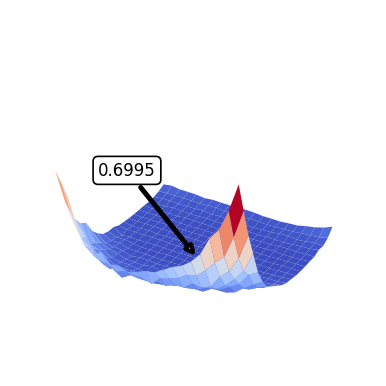}
    \subcaption{\textbf{T1wCE}}
\end{subfigure}
\end{minipage}
\begin{minipage}{0.24\textwidth}
\begin{subfigure}{\textwidth}
    \includegraphics[width=\textwidth]{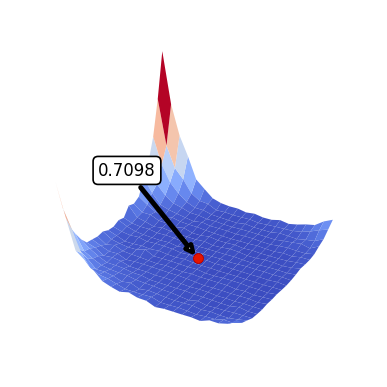}
    \subcaption{\textbf{FLAIR}}
\end{subfigure}
\end{minipage}
\begin{minipage}{0.24\textwidth}
\begin{subfigure}{\textwidth}
    \includegraphics[width=\textwidth]{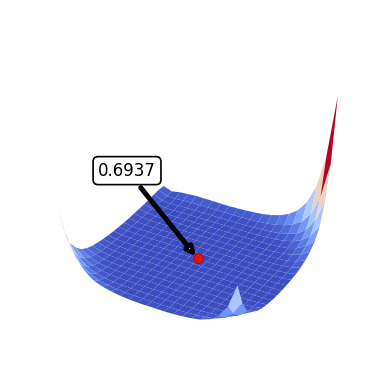}
    \subcaption{\textbf{Multi-Modal}}
\end{subfigure}
\end{minipage}
\caption{Binary cross-entropy loss landscapes of ResNet-10 model on the BRaTS2021 dataset.}
\label{data_sample2}
\end{figure*}

Another perspective to explore is the loss landscape of the trained models. We examine the curvature and the landscape of the loss function in Figures~\ref{data_sample} and~\ref{data_sample2} using the “filter normalization” technique \citep{li2018visualizing}. We plot the loss landscapes for ResNet-10 and ResNet-34 based on the BCE loss function shown in Equation~\ref{eq:BCE}. We choose the BCE loss as it converges better than other loss functions based on a previous study for the same task \citep{emchinov2022deep}. In all scenarios, we can observe from Figures~\ref{data_sample} and~\ref{data_sample2} that the loss landscapes have a flat curvature around an average value of 0.7 towards the end of training. To explain this behavior, Equation~\ref{eq:BCE} is solved under the assumption that both classes are almost equally represented in the training set, and the model is in a random state, i.e., both classes are predicted with a probability of 0.5. Solving Equation~\ref{eq:BCE} yields a loss of 0.69, which explains that these models are in a random state even after being trained for many epochs. On the other hand, the loss landscape of the BCE for the NoduleMNIST3D dataset shows a reduction in the loss (0.3086) compared to the random state of (0.69), suggesting that the model is able to learn representative features that split the two classes apart.

\begin{equation}\label{eq:BCE}
Loss = - \frac{1}{N} \sum_{i=1}^{N} y_i \times \log(p(y_i)) + (1-y_i) \times \log(1-p(y_i)) 
\end{equation}

\subsection{Limitations in Generalizability}
The disagreement between the cited articles is, in our opinion, due to several factors.
The lack of rigor in the validation process and the limited size of the datasets can result in an inflated performance. For instance, \cite{chang2018deep}, who achieved high performance accuracy, noted in their discussion that the lack of an independent dataset and a relatively small sample size ($N = 259$) could have impacted their ability to evaluate the generalization performance of their proposed model. Similarly, \cite{korfiatis2017residual} pointed out that they relied on data from a single source, which should be avoided to produce generalizable results. Likewise, \cite{yogananda2020mri}, who also claimed excellent precision, discussed potential data leakage problems in their earlier work.
It is important to note that their work has poor segmentation accuracy but excellent MGMT-promoter status classification accuracy. We attempted to replicate some of the earlier ideas and models which reported high performance, such as using segmentation masks to combine modalities for region-of-interests. However, those attempts failed to achieve high classification accuracy. One challenge we face is that most of the earlier cited work's source code is unavailable. 

The results of any study can be altered based on the development of novel datasets and methodologies. Thus, it might still be possible to predict the methylation status by combining other biomarkers or prognostic factors. Our study reminds the AI-centered scientific community about the importance of thorough and unbiased validation for future studies and clinical implementation. Our contribution encourages extending the research ground to investigate other prognostic biomarkers and raises ethical concerns by giving a clear example of the inability of the current deep learning models and datasets to diagnose the MGMT promoter from MRI scans.

\begin{figure}[!b]
\centering
\includegraphics[width=0.6\columnwidth]{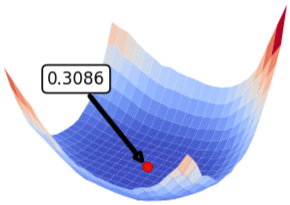}
\caption{Binary cross-entropy loss landscape of ResNet-10 model on the NoduleMNIST3D dataset.}
\label{NoduleLoss}
\end{figure}

\section{Recommendations}
There has been a recent uptick in the number of papers claiming that deep learning algorithms can perform as well as, if not better than, human oncologists in the diagnosis and treatment of cancer \citep{hosny2018artificial,vamathevan2019applications,bera2019artificial}. However, few systems have shown actual medicinal value \citep{nagendran2020artificial}. Similar to the study conducted in this work, other authors have found issues with the deep learning models focusing on the posture of patients, hospital instruments, marks on the skin, and hospital prevalence from the dataset to detect the disease \citep{narla2018automated, zech2018variable, winkler2019association}. This bias or over-reliance on specific features or variables can be problematic. If these systems are not properly evaluated, there is a risk that they will be based on flawed or biased assumptions and provide inaccurate or misleading results. This can lead to misdirection of further research, diminished credibility of research findings, and potentially even harmful to patients if these systems are used to influence treatment decisions \citep{topol2019high}. In order to avoid these negative consequences, it is crucial to develop and use standardized and rigorous evaluation protocols to assess the accuracy and reliability of deep learning systems for cancer diagnostics such as in the case of detection of MGMT promoter status. 

Hence we recommend the following suggestions for the development and evaluation process of deep learning models used in cancer diagnosis:

\begin{enumerate}
    \item \textit{Collect data from a diverse range of patients and tumors}: In order to develop a more diverse dataset for training and evaluating deep learning models, it is essential to collect data from a wide range of patients with different tumors properties \citep{noor2020can}. This can include data from patients of different ages, genders, races, and ethnicities and consists of tumors with different sizes, shapes, and characteristics \citep{gianfrancesco2018potential}. 
    By collecting such diverse datasets, it will be possible to train and evaluate deep learning models that are more robust and generalizable.
    \item \textit{Use external validation datasets}: In addition to collecting data from a diverse range of patients, it is also important to use external validation datasets to evaluate the performance of deep learning models. This can be achieved by evaluating the models on datasets from a cohort not part of the training dataset. By using external validation datasets, it will be possible to assess deep learning models' generalizability and robustness and identify any biases or limitations in their performance.
    \item \textit{Conduct explainability analyses}: Explainability analyses can also be used to evaluate the performance of deep learning models and identify any biases or limitations in their performance. These analyses can provide insights into the factors that are most important for the model's predictions and can help to identify any patterns or trends in the data that may be driving the model's decisions. By conducting explainability analyses, it will be possible to identify any potential biases or limitations in the model's performance and to take steps to correct or mitigate these issues. As demonstrated in this work with the help of Grad-CAM, occlusion sensitivity, and feature visualizations, we concluded that the models could not learn representative features.  
    \item \textit{Engage with clinicians and other stakeholders}: For deep learning models to be used effectively in a clinical setting, it is crucial to engage with clinicians and other stakeholders who will be using these models. This can involve engaging clinicians and other experts in developing and evaluating deep learning models, as well as soliciting their feedback and input on the design and implementation of these models. By engaging with clinicians and other stakeholders, it will be possible to develop deep learning models that are more closely aligned with the needs and goals of these users and can be more easily integrated into clinical practice.
    \item \textit{Use multi-modal data}: In order to develop and evaluate deep learning models for cancer diagnosis, it is vital to use a combination of data sources, including both structured and unstructured data. Structured data, such as electronic medical records, can provide valuable information about patients and their medical histories, while unstructured data, such as radiology images, can provide insights into the characteristics and appearance of tumors. Using a combination of data sources makes it possible to develop deep learning models that are more comprehensive and holistic and can provide a complete picture of the patient and their cancer.
    \item \textit{Develop standardized protocols and evaluation metrics}: To enable the comparison of deep learning models for cancer diagnosis, it is essential to develop standardized protocols and evaluation metrics. These protocols and metrics can provide a consistent and objective way to assess the performance of deep learning models. In addition, they can allow researchers and clinicians to compare different models' results and identify the most effective and reliable approaches.
    \item \textit{Reproducibility}:
    Deep learning models employed in cancer diagnosis must be reproducible, so researchers should be encouraged to share their code. Reproducibility refers to the ability of other researchers to acquire the same results using the same techniques and data, which is critical for determining the model's validity and reliability. In addition, reproducibility is essential for establishing trust in a study's findings and expanding scientific understanding through outcomes verification. Furthermore, repeatability is required for the therapeutic application of study findings and the generalization of results to a larger population. To summarize, the consistency of deep learning models employed in cancer diagnosis is critical for deep learning model validity, trust, verification, and clinical application.
    
\end{enumerate}

Deep learning systems have demonstrated excellent generalization performance on various tasks thanks to the ongoing research in the domain of generalization in deep learning, which utilizes certain design principles such as the recommendations above.

\section*{Conclusion}
% TO REVIEW
% Despite the extensive range of models that we tried and the models that show high performance in predicting MGMT status from previous research, none of them seem to be better than a random guess when it comes to predicting MGMT status. It appears that the model is assessing the tumor area. Still, it could not identify features that differentiate between the two classes based on our interpretation of Grad-Cam, t-SNE, PCA, and histograms. It seems that MRI images and MGMT status do not correlate. However, future research can focus on other biomarkers and perhaps establish a link between them.

The scientific investigation of deep learning approaches in medical diagnosis and prognosis is gaining popularity in the community; however, these studies' medical applications and effectiveness are limited in practice. In the case of MGMT promoter methylation status prediction, some performance discrepancy exists in the literature. This work examines a multitude of deep learning models to predict the MGMT status from MR scans from the largest cohort that is publicly available. We extensively study the models' capabilities to use the imaging data to predict the methylation status and test them using Grad-CAM, occlusion sensitivity analysis, feature visualizations, and training loss landscapes. We find no correlation between the imaging data and MGMT methylation status, concluding that the reliability of the deep learning approaches should be verified using external cohorts. We recommend a list of actions to take for future studies before considering medical applications.

% \section*{Acknowledgments}
% Acknowledgments should be inserted at the end of the paper, before the
% references, not as a footnote to the title. Use the unnumbered
% Acknowledgements Head style for the Acknowledgments heading.

\bibliographystyle{model2-names.bst}\biboptions{authoryear}
\bibliography{refs}

\end{document}